\documentclass[%
 reprint,
showpacs,
 amsmath,amssymb,
 aps,
]{revtex4-1}

\usepackage{fancyhdr}
\usepackage{graphicx}
\usepackage{dcolumn}
\usepackage{bm}

\usepackage[none]{hyphenat}

\begin{document}

\title{Second-Order Nonlinear Response in Centrosymmetric Hyperbolic Media.}%

\author{Evgenii E. Narimanov$^1$}
\affiliation{$^1$School of Electrical and Computer Engineering and Birck Nanotechnology Center, Purdue University, West Lafayette, Indiana 47907, USA}
\date{\today}

\begin{abstract}
We show that centrosymmetric natural hyperbolic media such as hexagonal boron nitride exhibit a bulk second-order nonlinear response, leading to second-harmonic and sum- and difference-frequency generation with efficiencies comparable to those of established nonlinear optical crystals such as potassium dihydrogen phosphate.
\end{abstract}

\maketitle

\section{Introduction}

Second-harmonic generation is conventionally associated with non-centrosymmetric media,\cite{Bloembergen}  as the electric-dipole contribution vanishes in systems with inversion symmetry.\cite{Boyd} At the same time, it has long been recognized that higher-order contributions involving field gradients can give rise to a second-order response even in centrosymmetric materials.\cite{Bloembergen,Shen}  These terms, however, have been treated as negligible corrections, as the characteristic scale of electromagnetic field variation in conventional systems is too large to produce a substantial effect.\cite{Bloembergen,Shen}   As a result, no regime has been identified in which such contributions yield a strong nonlinear response.

Such a regime does not arise in conventional electromagnetic settings, where outside the immediate near-field region the characteristic spatial variation of the field is fundamentally limited by the wavelength scale.\cite{BornWolf} This conclusion was established before the discovery of hyperbolic media,\cite{PN,hyperlens1,hyperlens2} a distinct class of materials whose anisotropic response supports propagation with arbitrarily large wave vectors and correspondingly strong spatial variation.\cite{PDOSprl,PDOSapl,ScienceTT} In natural hyperbolic materials such as hexagonal boron nitride,\cite{hBN1,DaiBasov2015} this behavior occurs in low-loss spectral bands and leads to deeply subwavelength field confinement.\cite{hBN2,hBN3,hBN4}

We uncover this regime when external fields generate strongly varying electromagnetic distributions within bulk hyperbolic media. In this setting, where the necessary subwavelength spatial structure of the field can be imposed, for example, by a subwavelength pattern at the interface or in the bulk, the second-order nonlinear response builds up in the bulk of a centrosymmetric hyperbolic medium. The resulting second-harmonic and sum- and difference-frequency generation is determined by the hyperbolic spatial variation scale and can reach efficiencies comparable to those of established nonlinear optical crystals such as potassium and ammonium dihydrogen phosphates.

\begin{figure}[htbp] 
   \centering
   \includegraphics[width=3.25in]{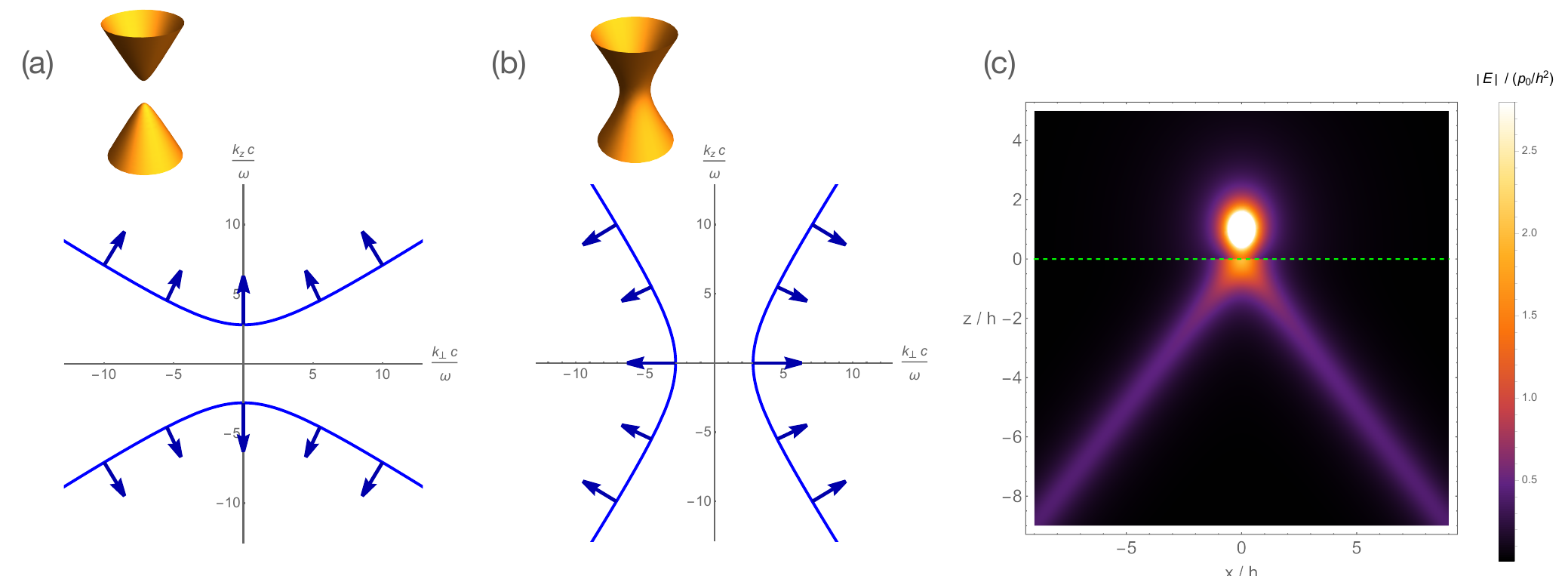} 
  \caption{Panels (a) and (b): Iso-frequency surfaces (insets)
and their cross-sections (main panels) in a uniaxial hyperbolic
medium, with $\epsilon_\perp > 0$, $\epsilon_z < 0$ in (a), and
$\epsilon_\perp < 0$, $\epsilon_z > 0$ in (b). Panel (c): Electric
field intensity (false color) generated by a subwavelength defect
at the interface with a hyperbolic medium, showing radiation
confined to two narrow sheets propagating at angles set by the
asymptotic directions of the iso-frequency surface. The field
distribution is calculated using material parameters of hexagonal
boron nitride \cite{hBN1,hBN2,hBNisotope} at the free-space
wavelength of $\lambda_0 \simeq 6.55 \ \mu{\rm m}.$}
   \label{fig:1}
\end{figure}

\section{The Mechanism}

While second-order nonlinear response is conventionally associated with the dipolar response to a uniform field in a non-centrosymmetric medium,\cite{Boyd} a more complete treatment reveals additional contributions.\cite{Bloembergen,Shen} In particular, spatial variation of the electromagnetic field gives rise to a term proportional to the product of the field and its gradient,\cite{Bloembergen,Shen}  providing a second-order response even in centrosymmetric systems, with the nonlinear polarization
\begin{eqnarray}
{ \bf P}^{(2)}\left(2 \omega\right) & = & {\bf \Lambda} \, {\nabla\bf E}\left(\omega\right)  \, {\bf E}\left(\omega\right)
\equiv {\bf \hat{e}}_\alpha \, {\Lambda}_{\alpha\beta\gamma\nu}  \frac{\partial E_\gamma}{\partial x_\beta} E_\nu , 
\end{eqnarray}
where ${\bf \hat{e}}_\alpha$ is the unit basis vector, and  the  nonlinear  susceptibility ${\bf \Lambda}$ of the centrosymmetric medium can be related to its microscopic  structure, that also defines its linear  response. In particular, using the anharmonic oscillator model  \cite{Bloembergen,Garrett1968} of the material response, we obtain
\begin{eqnarray}
{\Lambda}_{\alpha\beta\gamma\nu} & = & \frac{ \chi_{\alpha\gamma}(2\omega)\, \chi_{\beta\nu}(\omega)}{Ne}, 
\end{eqnarray}
where $\chi_{\alpha\beta}$ is the frequency-dependent linear-response susceptibility, $N$ is the atomic number density in the medium and $e$ is the electron charge. Note that this mechanism does not rely on any anharmonicity of the crystalline potential, and is entirely governed by the linear response of the medium.

In conventional electromagnetic settings, however, this contribution is strongly suppressed,\cite{Bloembergen,Shen}   since spatial variations of the field outside the immediate near-field  are limited by the wavelength scale.\cite{BornWolf}  The associated field gradients ${\bf \nabla E}$  therefore remain small, and the resulting nonlinear response is correspondingly weak,\cite{Bloembergen,Shen} being controlled by the inherently small parameter given by the ratio of the characteristic spatial variation scale to the free-space wavelength, so that such terms can be treated as negligible corrections.\cite{Boyd}

However, this conventional picture is profoundly recast in hyperbolic media,\cite{PN}  a recently discovered class of materials whose unusual electromagnetic response has enabled striking effects such as imaging beyond the diffraction limit.\cite{hyperlens1,hyperlens2}  In these systems, the wavelength no longer sets the fundamental scale of spatial variation: the anisotropic dispersion supports propagation with arbitrarily large wave vectors,\cite{PN} and the gradients that govern this contribution are no longer constrained to be small.

As a result, the suppression that renders this contribution negligible in conventional media \cite{Boyd} is removed, and the  field-gradient term becomes a substantial source of second-order response.
This behavior can be traced to the distinctive electromagnetic dispersion of hyperbolic media. Unlike conventional materials, where iso-frequency surfaces are closed and the wave vector is bounded by its free-space value,\cite{BornWolf} hyperbolic media support open dispersion surfaces that extend to arbitrarily large wave vectors.\cite{PDOSprl,PDOSapl,ScienceTT}

Hyperbolic media form a class of strongly anisotropic materials in which the principal components of the dielectric permittivity tensor have opposite signs. This unusual response leads to open iso-frequency surfaces \cite{ScienceTT} and underlies the distinctive propagation properties discussed above. Such behavior can arise, for example, from the anisotropic motion of free carriers \cite{ScienceTT} or from phonon--polariton resonances in anisotropic crystals.\cite{hBN1,DaiBasov2015,hBN2,hBN3,hBN4}

\begin{figure*}[htbp] 
   \centering
   \includegraphics[width=6.5in]{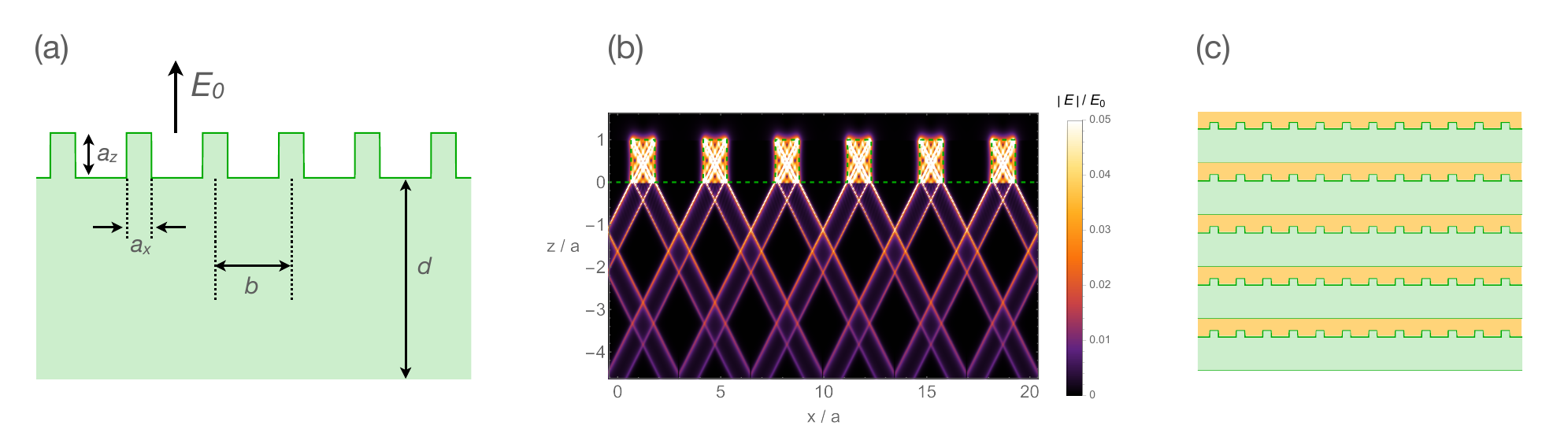} 
\caption{(a) The array of subwavelength defects separated by distance $b$ at the interface with a hyperbolic medium of thickness $d$. (b) Induced electric field distribution (not including the 
``incident'' field ${\bf E}_0$ -- see panel (a))  showing the superposition of the fields generated by individual defects and the resulting subwavelength spatial structure in hexagonal boron nitride at the at the free-space
wavelength of $\lambda_0 \simeq 6.55 \ \mu{\rm m}$. The second-order response is determined by the corresponding nonlinear polarization averaged over the layer with respect to the external field $E_0$. (c)Bulk implementation of the proposed mechanism, forming a nonlinear medium rather than a surface layer. The structure essentially represents a hypercrystal with embedded subwavelength coupling elements.}
   \label{fig:2}
\end{figure*}

In a uniaxial hyperbolic medium with the $\hat{\bf z}$ axis chosen parallel to the symmetry axis, so that $\epsilon_x = \epsilon_y = \epsilon_\perp$, with $k_\perp \equiv \sqrt{k_x^2 + k_y^2}$ the resulting iso-frequency surface for propagating $p$ (TM)-polarized electromagnetic waves is defined by
\begin{eqnarray}
\frac{k_z^2}{\epsilon_\perp} + \frac{k_\perp^2}{\epsilon_z} & = & \frac{\omega^2}{c^2},
\label{eq:hyperbola}
\end{eqnarray}
and is described by a hyperboloid of revolution (see Fig. \ref{fig:1}(a),(b)). The propagation direction is then given (in the limit of small loss) by the group velocity ${\bf v}g = \partial\omega_{\bf k}/\partial{\bf k}$ normal to the iso-frequency surface. For large wavenumbers, which dominate the phase space of the hyperbolic medium (see Fig. \ref{fig:1}(a),(b)), the propagation direction is therefore set by the angle
\begin{eqnarray}
\vartheta_\omega \equiv \arctan {\rm Re} 
\left[ \sqrt{ - \frac{\epsilon_\perp\left(\omega\right)}{\epsilon_z\left(\omega\right)}} \, \right]
\end{eqnarray}
with respect to the symmetry axis $\hat{\bf z}$ for large wave vectors. This behavior is directly seen in Fig. \ref{fig:1}(c), where the radiation scattered by the subwavelength defect into the hyperbolic medium is confined to two narrow sheets propagating at these angles, providing the subwavelength field structure that can drive a second-order nonlinear response in a centro-symmetric medium. As the hyperbolic medium supports propagating modes with large wavenumbers, for small material loss the ``light sheets'' remain tightly collimated, with a transverse width comparable to the size of the defect, allowing the subwavelength field structure to persist deep into the bulk of the medium rather than being confined to the near-field region. Note that the field distribution shown is obtained using the actual material parameters in hexagonal boron nitride ($h$BN), without recourse to any lossless or idealized approximation.

The subwavelength defect at the interface therefore acts as an effective dipolar source for the high-wavenumber modes of the hyperbolic medium.
For this configuration, a straightforward calculation (see Appendix A) yields the electric field in the hyperbolic medium
\begin{eqnarray}
{\bf E} & = & - \nabla \phi,
\label{eq:E1}
\end{eqnarray}
where the scalar potential
\begin{eqnarray}
\phi & = & - \frac{\epsilon_0}{\epsilon_z}  E_0  z  \left[ 
1 -  \frac{  \alpha_{\rm eff} A }{1 - \frac{i \, \epsilon_0}{ \sqrt{-\epsilon_\perp \epsilon_z} }  } \, 
\frac{2 \epsilon_z }{ \epsilon_z \, x^2  +  \epsilon_\perp \, z^2 + i z  h \delta  }  \right], \ \ \ \ 
\label{eq:phi1}
\end{eqnarray}
where $E_0 \hat{\bf z}$ is the external electric field (see Fig. \ref{fig:1}), $\epsilon_0$ is the dielectric permittivity of the medium above the hyperbolic layer (see Fig. \ref{fig:1}(c)), $A$ is the cross-section area of the defect,  
and $\alpha_{\rm eff} = {\cal O}\left(1\right)$ is its effective dimensionless polarizability that depends on the defect
composition and geometry -- see Appendix B, $h$ is the distance from the hyperbolic interface to the center of the defect,  and
$
\delta  =  2 \epsilon_\perp / \sqrt{-\epsilon_\perp \epsilon_z} .
$ This solution explicitly shows the subwavelength spatial structure of the field generated by the subwavelength defect, which then drives a second-order nonlinear response in a centrosymmetric medium.

Note that a finite material loss, while small in natural hyperbolic media such as $h$-BN, leads to a progressive suppression of the large-wavevector components, with attenuation increasing with wavevector magnitude, that define the narrow light sheets. As a result, with increasing distance from the defect the peak intensity decreases and the hyperbolic light-sheets broaden. For the geometry of Fig. \ref{fig:1}(c), the transverse width of the hyperbolic light-sheet evolves as
\begin{eqnarray}
w & = & {h}{} + \left| {\rm Im} \left[ \sqrt{ - \frac{\epsilon_\perp}{\epsilon_z}}  \, \right] z \right|,
\end{eqnarray}
so that the spatial scale of the field increases with propagation distance, leading to a corresponding reduction of the field-gradient nonlinear response.
While material loss limits the propagation length of large‑wavevector modes, in natural hyperbolic media such as hexagonal boron nitride this length remains sufficient for the nonlinear response to build up over a bulk region.

With the thickness of the hyperbolic medium chosen to fully capture the light sheets before they decay, reflections from the substrate beneath the hyperbolic layer can be neglected, so that the fields generated by different defects do not couple. For the array of defects separated by distance $b$ in Fig. \ref{fig:2}(a), the total field is therefore given by the superposition of the individual defect contributions,
\begin{eqnarray}
{\bf E}({\bf r}) & = & \sum_n {\bf E}_1\left({\bf r} - \hat{\bf x} \, b \, n\right),
\end{eqnarray}
where ${\bf E}_1$ is the field of a single defect defined in Eqns. (\ref{eq:E1}),(\ref{eq:phi1}).

The characteristic dimensions of the structure are constrained by the physics of the coupling and propagation in the hyperbolic medium (see Fig. \ref{fig:2}). Efficient excitation of the large-wavevector modes requires the defect size $a$ to remain deeply subwavelength, while a small spacing $b$ is needed to ensure a sufficiently dense array of sources and the formation of multiple light sheets. The thickness $d$ of the hyperbolic layer is in turn limited by material loss: even in high-quality crystals, including isotopically purified samples, where the quality factors \cite{EN2025} $Q = 1/{\rm Im}\,\sqrt{-\epsilon_\perp/\epsilon_z}$ remain on the order of $\sim 10^2$ \cite{hBNisotope}, so that the light-sheets broaden with propagation (Eq.~(\ref{eq:phi1})) as $w \sim a + z/Q$. With the transverse scale $w \lesssim 10 a$, the propagation distance is $z \sim 100 \, a$, corresponding to $d \lesssim 1~\mu{\rm m}$ for representative parameters. In low-loss hyperbolic materials \cite{hBNisotope} such as hexagonal boron nitride, the hyperbolic bands occur at wavelengths $\sim 10~\mu{\rm m}$ or above, so that the entire structure remains deeply subwavelength on the scale of the external field.

At the same time, the nonlinear polarization is generated by strongly nonuniform local fields associated with the hyperbolic light sheets, whose spatial scale evolves with propagation. Thus, the response of the system is intrinsically inhomogeneous on the microscopic level. However, since the entire structure remains deeply subwavelength on the scale of the external electromagnetic field, this internal structure cannot be resolved by the incident radiation. The relevant observable is therefore the polarization averaged over the layer.

In this regime, the nonlinear response of the structured hyperbolic medium is characterized by an effective second-order susceptibility, defined with respect to the external driving field through the relation
\begin{eqnarray}
P^{(2)} & = & \chi^{(2)}_{\rm eff} \, E_0^2,
\end{eqnarray}
where $P^{(2)}$ is the nonlinear polarization averaged over the layer and $E_0$ is the amplitude of the external field (see Fig. \ref{fig:2}), that arises from the subwavelength field structure generated within the hyperbolic medium.

Using Eqn. (\ref{eq:phi1}), we obtain
\begin{eqnarray}
\left( \chi^{(2)}_{\mathrm{eff}} \right)_{zz}
& = &
\frac{\epsilon_\perp(\omega)}{ \epsilon_z(\omega)}
\frac{\chi_{zz}(\omega)\, \chi_{zz}(2\omega) \, \alpha_{\rm eff}^2(\omega) }{(1 + i \sqrt{-\epsilon_z(\omega)\epsilon_\perp(\omega)}/\epsilon_0)^2}
\frac{\pi  a}{e N b d} \ \ \ \ \ \nonumber \\
& = & {\cal O}\left( \frac{ a}{e N b d} \right), 
\label{eq:chi2}
\end{eqnarray}
where the hyperbolic layer thickness $d  \gtrsim  Q {a} $, and $N$ is the atom density of the hyperbolic material.

With $a \sim 10~{\rm nm}$, $h \sim a$, $b \sim 5a$, and $d \sim 10a$, we find
\begin{eqnarray}
\chi^{(2)}_{\rm eff} & \sim & 10^{-9}\ {\rm esu}.
\end{eqnarray}
For comparison, in ADP $\chi^{(2)}_{xyz}=\chi^{(2)}_{yxz}\approx 2.4\times 10^{-9}\ {\rm esu}$ and $\chi^{(2)}_{zzz}\approx 2.4\times 10^{-9}\ {\rm esu}$, while in KDP $\chi^{(2)}_{xyz}=\chi^{(2)}_{yxz}\approx 2.4\times 10^{-9}\ {\rm esu}$ and $\chi^{(2)}_{zzz}\approx 2.2\times 10^{-9}\ {\rm esu}$. Thus, the effective second-order response of the centrosymmetric hyperbolic medium is on the same scale as in these widely used non-centrosymmetric nonlinear crystals.

The proposed mechanism can be implemented in a macroscopically thick geometry (Fig.~\ref{fig:2}(c)), forming a bulk nonlinear medium rather than a structure with a surface‑confined response. Such configurations represent a particular realization of hypercrystals, originally proposed in Ref.~\cite{PRX}, and subsequently demonstrated experimentally in multiple geometries, including structures with embedded gain.\cite{Vinod,Vinod2D} We stress that the nonlinear mechanism uncovered in the present work applies to all hypercrystal designs and geometries.~\cite{Vinod}

We emphasize that the comparison above is made at the level of the effective nonlinear susceptibility, which directly carries over to the bulk implementation. As in conventional nonlinear optical materials such as KDP, enhanced nonlinear output through phase matching can be achieved through appropriate system design in structures based on the present mechanism.

\appendix

\section{Line Dipole at a Hyperbolic Interface.}

Here we consider a planar interface between an isotropic dielectric (permittivity $\varepsilon_0$) and a uniaxial hyperbolic medium (permittivities $\varepsilon_z$ and $\varepsilon_x = \varepsilon_y = \varepsilon_\perp$) occupying the half-space $z<0$. The source is a {\it line dipole}, uniform along one transverse direction (taken to be $\hat{\mathbf{y}}$), and located at a distance $h$ from the interface, with the  dipole moment oriented along $\hat{\mathbf{z}}$:
\begin{eqnarray}
\mathbf{p} & = & p_z \hat{\mathbf{z}}.
\end{eqnarray}
In the quasi-static regime relevant to deeply subwavelength structure, the field is described in terms of a scalar potential $\phi$, with $\mathbf{E} = -\nabla \phi$. The resulting solution can then be expressed in the form
\begin{eqnarray}
\phi(\mathbf{r}) & = & - \frac{p_z \left(z - h\right) \, \theta\left(z\right) }{(z-h)^2 + x^2} 
+  \int_0^\infty  dq \, \cos\left( q x\right)   \nonumber \\ 
& \times & \left(  r_q \,  e^{- q z} \, \theta\left(z\right)  + t_q \, e^{- i  k_q z }  \, \theta\left(-z\right)  \right), 
\end{eqnarray}
where $\theta\left(x\right)$ is the Heaviside function, $r_q$ and $t_q$ are the reflection and transmission coefficients, and
\begin{eqnarray}
k_q & = & \sqrt{ - \frac{\varepsilon_\perp }{\varepsilon_z}} \, {\rm sign} \left[ {\rm Re}\, \epsilon_\perp \right] q.
\end{eqnarray}

Matching the scalar potential and the resulting normal component of the displacement vector ${\bf D} = \epsilon \, {\bf E} $ at the interface, we obtain
\begin{eqnarray}
\phi(\mathbf{r}) & = & - \frac{p_z \left(z - h\right) \, \theta\left(z\right) }{(z-h)^2 + x^2} \nonumber \\
& - &  
\frac{ {\epsilon_0} + i \, \sqrt{-\epsilon_z \epsilon_\perp} }
{  {\epsilon_0} - i \, \sqrt{-\epsilon_z \epsilon_\perp}} \cdot 
\frac{p_z \, \left( z+h\right) \, \theta\left(z\right)}{(z+h)^2 + x^2} \nonumber \\
& - &  \frac{2  \epsilon_0 }{ \epsilon_0 + i \, \sqrt{-\epsilon_z \epsilon_\perp} }
\nonumber \\
& \times & 
\frac{p_z \left( \epsilon_z  h - i  \, \sqrt{-\epsilon_\perp \epsilon_z} \, z \right) \, \theta\left(-z\right)}
{
\epsilon_\perp z^2 + \epsilon_z \left( x^2 + h^2\right) 
+ 2 \, i \, \sqrt{-\epsilon_\perp \epsilon_z} \, z  h 
}.
\label{eq:phiA1}
\end{eqnarray}
Note that for $z > 0$ the scalar potential (\ref{eq:phiA1}) corresponds to the superposition of the field of the original dipole and its ``image'' 
\begin{eqnarray}
p_z' = p_z \, \frac{ {\epsilon_0} + i \, \sqrt{-\epsilon_z \epsilon_\perp} }
{  {\epsilon_0} - i \, \sqrt{-\epsilon_z \epsilon_\perp}}. 
\end{eqnarray}
located at $z' = - h$.

For $z < 0$, our solution (\ref{eq:phiA1})  captures the strongly anisotropic spatial structure associated with hyperbolic dispersion, including the emergence of highly directional field distributions that form the basis of the nonlinear response discussed in the main text.

Introducing the effective dimensionless polarizability $\alpha_{\rm eff}$ (see Appendix B), which relates the dipole moment to its cross-section area $A$ and the external field $E_0 \hat{\bf z}$ (see Fig. 1(c)) via
\begin{eqnarray}
p_z & = & \alpha_{\rm eff}(\omega)\, A \, E_0,
\label{eq:alpha}
\end{eqnarray}
for $\left|x\right|, \,  \left|z\right| \gg h$, which excludes the immediate near field of the source dipole, we immediately obtain Eqn. (6) of the main text.

\begin{figure}[htbp] 
   \centering
   \includegraphics[width=3.3in]{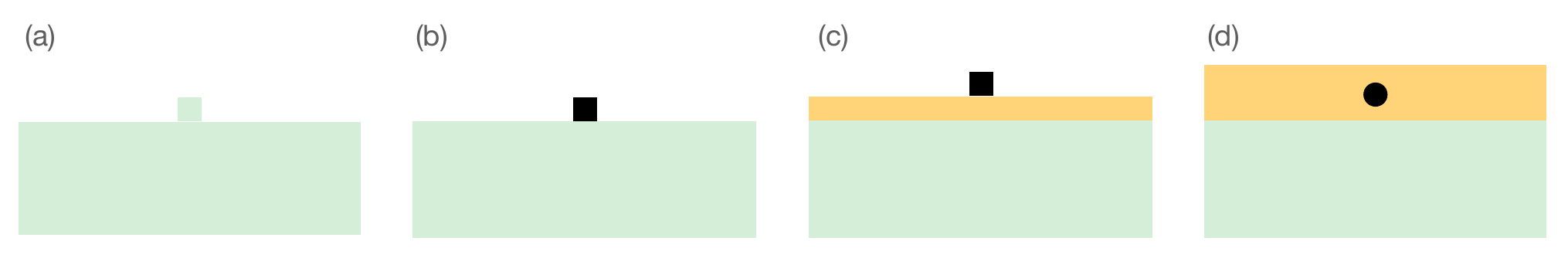} 
   \caption{Representative realizations of the subwavelength defect coupling the external field to high-wavenumber modes in the hyperbolic medium: (a) ridge etched into the material, (b,c) high-index nanowire deposited on the surface,  with (c) and without (b) a spacer layer, and (d) buried nanowire within a dielectric  near the hyperbolic interface.}
   \label{fig:A1}
\end{figure}

\section{}

Depending on the practical constraints for device fabrication (see Fig. \ref{fig:A1}), the defects that couple the external electric field to the high-wavenumber fields in the hyperbolic medium can be formed by either a ridge/trench etched into the material (Fig. \ref{fig:A1}(a)), or a high-refractive-index nanowire deposited onto the surface of the hyperbolic medium, with (Fig. \ref{fig:A1}(b)) or without an additional spacer layer, or with the line defect surrounded by the spacer material (Fig. \ref{fig:A1}(c)). In the latter case, the increased fabrication complexity is motivated by the use of a nanowire supporting a plasmon resonance in the surrounding dielectric environment, with the corresponding enhancement of the scattering efficiency.

In the ridge geometry of Fig.  \ref{fig:A1}(a), for the scalar potential in the hyperbolic medium $\left|x\right|, \left|z\right| \gg a_x, a_z$ we obtain
\begin{eqnarray}
\phi & = & \frac{1}{\pi^2}
\left( 1 - \frac{\epsilon_0}{\epsilon_z} \right) \left( 1 - \frac{\epsilon_z} {\epsilon_0} \frac{a_x}{a_z} \right) \sqrt{-\frac{\epsilon_\perp}{\epsilon_z}} \nonumber \\
& \times &  \frac{a_x a_z E_0}{\epsilon_z x^2 + \epsilon_\perp z^2}  , 
\end{eqnarray}
where $a_x$ and $a_z$ are respectively the horizontal and the vertical dimensions of the ridge cross-section (see Fig. \ref{fig:A1}(a)). Therefore, in this case the effective polarizability defined by Eqn. (\ref{eq:alpha}), is given by
\begin{eqnarray}
 \alpha_{\rm eff} & = &  \frac{1}{\pi^2}
\left( 1 - \frac{\epsilon_0}{\epsilon_z} \right) \left( 1 - \frac{\epsilon_z} {\epsilon_0} \frac{a_x}{a_z} \right) \sqrt{-\frac{\epsilon_\perp}{\epsilon_z}}.
\end{eqnarray}

On the other hand, in the ``buried wire'' case of Fig. \ref{fig:A1}(c), we obtain
\begin{eqnarray}
 \alpha_{\rm eff} & = &  \frac{1}{4} 
 \frac{\epsilon_w - \epsilon_0}{\epsilon_w + \epsilon_0 - \left( \epsilon_w - \epsilon_0 \right) \left( \frac{a}{4 h}\right)^2 
  \frac{\epsilon_0 -  i \sqrt{ - \epsilon_z \epsilon_\perp}}{\epsilon_0 + i \sqrt{ - \epsilon_z \epsilon_\perp} }},
\end{eqnarray}
where $a$ is the diameter of the nanowire, $h$ is the  distance from its center and the hyperbolic medium, and 
$\epsilon_w$ is the dielectric permittivity of the nanowire.

\end{document}